# Tunneling study of cavity grade Nb: possible magnetic scattering at the surface

T. Proslier[1,2], J. F. Zasadzinski[1], L. Cooley[3], C. Antoine[4], J. Moore[2], M. Pellin[2], J.Norem[2], K.E. Gray[2]

[1]*Physics Division, Illinois Institute of Technology, Chicago, Illinois 60616*
[2]*MSD/HEP Divisions, Argonne National Laboratory, Argonne, Illinois 60439*
[3]*Technical Division, Fermi National Accelerator Laboratory, Batavia, Illinois 60510*
[4]*Commissariat à l'énergie atomique, Centre d'étude de Saclay F-91191 Gif-sur-Yvette*

Tunneling spectroscopy was performed on Nb pieces prepared by the same processes used to etch and clean superconducting radio frequency (SRF) cavities. Air exposed, electropolished Nb exhibited a surface superconducting gap $\Delta=1.55$ meV, characteristic of clean, bulk Nb. However the tunneling density of states (DOS) was broadened significantly. The Nb pieces treated with the same mild baking used to improve the Q-slope in SRF cavities, reveal a sharper DOS. Good fits to the DOS were obtained using Shiba theory, suggesting that magnetic scattering of quasiparticles is the origin of the gapless surface superconductivity and a heretofore unrecognized contributor to the Q-slope problem of Nb SRF cavities.

Elemental niobium (Nb) is the material of choice in two of the most important commercial superconducting devices: *i)* the Josephson tunnel junction (JTJ)[1] and *ii)* the superconducting radio frequency (SRF) cavity. The complex surface oxides of air-exposed Nb[2] have played an important role in the development of both devices over the past 30 years. For JTJs the deleterious effects of the surface oxides (reduced gaps and Josephson currents, large sub-gap quasiparticle currents, etc.) were eliminated by introducing an ultra-thin capping layer of Al, a technique first successfully established on Nb foils[3]. Prevention of oxygen exposure to the underlying Nb is the key to the construction of JTJ devices such as mixers and analog-digital converters, all of which now utilize Nb/Al bi-layer technology[4].

For SRF cavities the Nb oxide layers are relevant because they occupy a significant fraction of the region where electric and magnetic fields **E** and **B** are confined, within one magnetic penetration depth (~45 nm) from the surface. The reduction of the SRF cavity quality factor Q with increasing **E** field, (Q-slope problem) and its mitigation by a mild annealing procedure[5] (baking effect), are not well understood. There is evidence the baking effect is related to a decrease of $Nb_2O_5$ and an increase of $NbO_2$ from surface probes such as X-ray photoemission spectroscopy measurements applied before and after the baking[6]. However, the particular mechanism by which oxygen affects Q-slope is still elusive.

*Corresponding Author: prolier@iit.edu*

Here we report tunneling measurements on cavity-grade Nb that directly probe the surface superconductivity. Our results may provide new insights into the Q-slope problem and the baking effect. Air exposed, electropolished samples reveal a surface gap parameter characteristic of clean, bulk Nb ($\Delta=1.55$ meV). However the tunneling density of states (DOS) is considerably broadened, exhibiting >30% zero-bias conductance. Samples treated using the same mild baking step that reduces the Q-slope (e.g., 120°C for 24h - 48h) show much sharper DOS and reduced zero-bias conductance. These results are interpreted as indicating that magnetic scattering of quasiparticles, likely from the Nb oxide layers, produces a gapless superconducting surface layer that contributes to RF dissipation.

A pristine Nb surface exposed to air develops a complex set of oxides including NbO, $NbO_2$ and $Nb_2O_5$. Each of these oxides is thermodynamically stable with substantial off-stoichiometry. The topmost $Nb_2O_5$ layer is an ordinary band insulator with an energy gap >4 eV. Perhaps most significant is that sub-stoichiometric $Nb_2O_5$ (i.e. oxygen vacancies) develops magnetic moments[7] a property that has been generally ignored in SRF cavity development. Below $Nb_2O_5$ is a more complex Peierls semiconductor[8], $NbO_2$, with a band gap ~0.1 eV. Next, NbO is a normal metal ($T_c$=1.3 K) that can lead to proximity effects with the adjacent Nb. Finally, the solubility of O in Nb is rather large (up to 22%), so dissolved O in the Nb region closest to the oxides can also produce a layer of reduced superconductivity (so-called poisoned layer) since the $T_c$ of Nb is initially reduced by 1 K for each atomic percent of oxygen[9].

Tunnel junctions were formed using a mechanical contact between the Nb sample and an Au tip attached to a differential micrometer screw as described elsewhere[10]. Pieces of monocrystalline Nb (110), with a residual resistivity ratio (RRR) greater than 260, were cut from a larger sheet used to construct SRF cavities. The pieces were electropolished, cleaned with deionized purified water and dried in air in a manner similar to that done on cavities. One of the pieces was given a mild bake either in air or in vacuum at 125 C for 24 hours and then remeasured. Figure 1 shows a set of ten rescaled and shifted dynamic conductance curves (dI/dV vs. V) on the unbaked Nb piece taken at T=1.7 K. The junctions were formed independently in a single run by completely retracting the Au tip (~1 mm) and then re-advancing the tip until a tunnel current was observed. In this way, these junctions probe different regions over an area estimated to be greater than 10 μm$^2$. Each of the junctions exhibits a single gap-like feature close to the Nb bulk value (Δ=1.53-1.55 meV), but the significant broadening and large sub-gap conductance (of >30% of the normal state value, inferred from the flat region at higher bias voltages) suggest gaplessness. A similar set of ten junctions was obtained on the vacuum baked Nb piece. Both data sets were then rescaled and averaged, with the results shown in Fig.2. Samples baked in air show similar behavior as vacuum baked.

It is found, reproducibly, that baked Nb samples exhibit much sharper gap features in the conductance, characterized by a larger peak height to background and a smaller sub-gap conductance, than the unbaked Nb. This result immediately suggests a new interpretation of the baking effect: it reduces the normal, quasiparticle states inside the superconducting gap that give rise to dissipation and lower Q.

Prior to discussing the fits of the normalized data it is worth noting that there are numerous processes that could give rise to sub-gap conductance and therefore apparent gaplessness as measured by tunneling. These include proximity effects, severely degraded superconducting surfaces, and non-tunneling conductance processes (e.g. ohmic channels, hopping conductance, etc). However, a key feature of the data of Figs. 1 and 2 is a considerable peak shift toward higher voltages compared to the expectations of the Bardeen-Cooper-Schrieffer (BCS) DOS with Δ = 1.55 meV: this narrows the range of likely explanations. This peak shift is seen more clearly in Fig. 3 where the baked sample conductance is compared directly to a BCS conductance at 1.7 K. The inset of Fig. 3 shows that the conductance peak of the unbaked sample is beyond 2.0 meV. None of the above processes that produce sub-gap quasiparticle states would produce a concomitant shifting of the peak to higher voltages.

If anything, proximity effects and degraded layers lead to reduced gaps and would shift the peak to *lower* voltages. Furthermore, additional conductance channels from inelastic or resonant tunneling[11] would simply add to the overall measured conductance and not shift the peaks. The only mechanisms we are aware of that shift the conductance peak to higher voltages are pair-breaking phenomena such as those caused by strong-coupling effects[12] (so-called Dynes lifetime effect) or scattering from magnetic impurities[13].

We stress that the fundamental process behind the Dynes effect should not be relevant at $1.7K \ll T_C$ and furthermore the data are poorly fit by this expression as shown in Fig. 3. Also, the Dynes effect gives a more rounded feature near zero bias than the V-shaped data.

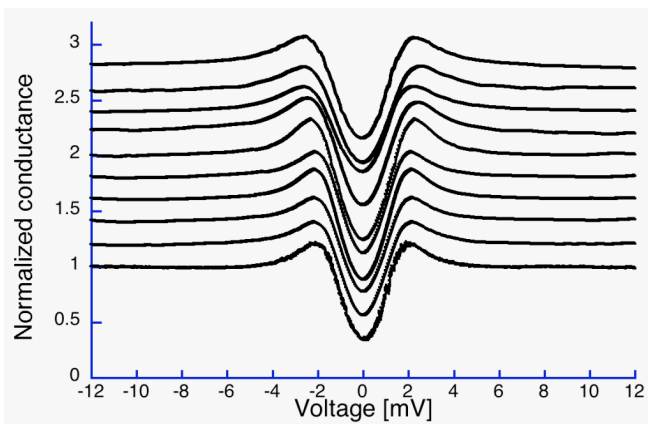

FIG. 1. (Color online) 10 typical conductance curves measured at 1.7 K on an EP unbaked Nb sample, shifted for more clarity.

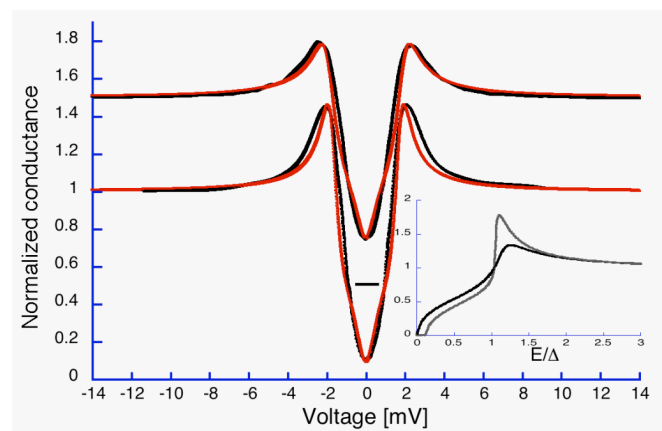

FIG. 2. (Color online) In black, average conductance curves of unbaked (top) and vacuum baked (bottom) Nb samples, shifted for more clarity. In red, fits using Shiba theory: Unbaked fit Δ=1.55 meV, α=0.32, ε=0.62, Baked fit: Δ=1.55 meV, α=0.2, ε=0.62. Inset: corresponding Shiba DOS

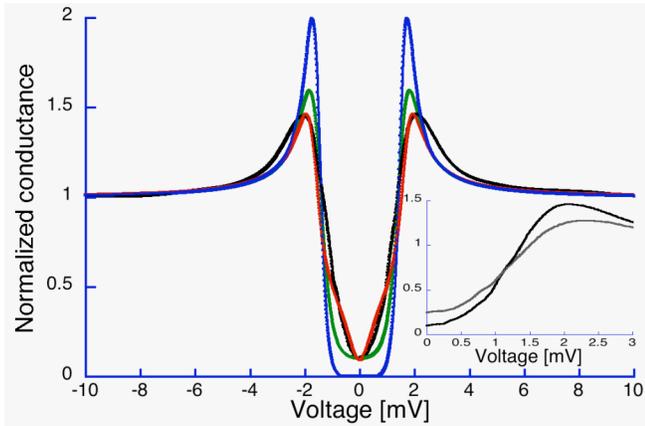

FIG. 3. (Color online) In black experimental baked conductance curve, in red fit using Shiba theory, green fit using the Dynes model ($\Gamma$=0.22 meV, $\Delta$=1.55 meV) and in blue the BCS conductance ($\Delta$=1.55 meV). Inset: peak voltage region of baked, unbaked average conductance curves.

This leads us to interpret the data in terms of magnetic scattering. Recognizing that sub-stoichiometric $Nb_2O_5$ will produce unscreened *d*-band magnetic moments we consult Shiba theory[13] that treats magnetic impurities in the strong scattering limit. Fits to the Shiba theory of the averaged rescaled conductances for baked and unbaked Nb are shown in Fig. 1 along with their corresponding DOS shown in Fig. 1 inset. These fits are best for both baked and unbaked data when the bulk superconducting energy gap parameter, $\Delta$=1.55 meV, is used. Considering the possibilities of proximity effects and a poisoned layer, this is an impressive confirmation of our premise.

In the Shiba theory, a band of quasiparticle states centered at an energy ε develops inside the superconducting gap. The scattering rate of all quasiparticles, $\Gamma_S = \alpha\Delta$ where α is the inelastic scattering parameter, broadens the band and shifts the DOS peak to an energy E>$\Delta$. Note that the Shiba model gives a more V-shaped conductance near zero bias in better agreement with the data. While the overall agreement with the experimental data is very good, there is still the issue of how superconducting pairs in Nb are scattered by magnetic impurities in a nearby oxide layer.

The fact that the Nb oxide layers are not simple planar surfaces, but rather jagged with occasional inclusions and precipitates makes the magnetic scattering picture more plausible. Furthermore, the reduction in magnetic scattering with mild baking does not require diffusion of oxygen over hundreds of angstroms, but instead could be brought about by a local rearrangement of oxygen: in particular, growth of a thicker[14], more stable, stoichiometric $NbO_2$ insulator, acting as a protective non-magnetic oxide. The $NbO_2$ would then prevent the adjacent Nb quasiparticles from being scattered by magnetic impurities present in the $Nb_2O_5$, located on top of the $NbO_2$. This hypothesis might explain why the improved performances of baked cavities are neither affected by an HF rinse (removal of the $Nb_2O_5$) nor by a thermal treatment in air or in vacuum[4].

In summary, tunneling measurements on SRF cavity grade Nb have provided new insights into the mechanism of the Q-slope problem and baking effect. The tunneling conductances reveal a gap parameter close to the optimal bulk value but with a broadened conductance for baked and unbaked Nb, indicating a pair-breaking process. This suggests that there is a contribution to RF dissipation arising from normal quasiparticle states inside the superconducting gap in a thin surface layer at the interface with the Nb oxide layers. A mild baking effect like that given to SRF cavities sharpens the conductance in the gap region and reduces the sub-gap quasiparticle density of states, thus remedying the dissipative effect. The only identifiable process for these effects is magnetic scattering, which is certainly plausible given the magnetic properties of reduced Nb oxides.

This work was supported by the U.S. DOE under contract DE-AC02-06CH11357.